\documentclass[conference]{IEEEtran}
\usepackage{cite}
\usepackage{amsmath,amssymb,amsfonts}
\usepackage{graphicx}
\usepackage{mathptmx}
\usepackage{xcolor}
\usepackage{subfigure}
\usepackage{booktabs}

\begin{document}

\title{TCAD Simulation of Novel Multi-Spacer HK/MG 28 nm Planar MOSFET for Sub-threshold Swing and DIBL Optimization}



\title{TCAD Simulation of Novel Multi-Spacer HK/MG 28 nm Planar MOSFET for Sub-threshold Swing and DIBL Optimization}

\author{\IEEEauthorblockN{Zhentao Xiao$^{1,2}$, Yihao Zheng$^1$, Zonghao Zhang$^{1,2}$, Jinhong Shi$^{1,2}$, Chenxing Wang$^{1,2}$, \
Yunteng Jiang$^{1,3}$, \\Haimeng Huang$^{1,*}$, Aynul Islam$^{1,2}$, Hongqiang Yang$^1$}
\IEEEauthorblockA{
$^1$University of Electronic Science and Technology of China, Chengdu, 610054, China\\
$^2$University of Glasgow, Glasgow, G12 8QQ, United Kingdom\\
$^3$Nanyang Technological University, 639798, Singapore\\
Email: hmhuang@uestc.edu.cn}}

\maketitle

\begin{abstract}
This study optimizes 28 nm planar MOSFET technology to reduce device leakage current and enhance switching speed. The specific aims are to decrease subthreshold swing (S.S.) and mitigate drain-induced barrier lowering (DIBL) effect. Silvaco TCAD software is used for process (Athena) and device (Atlas) simulations. We implemented our device (planar 28 nm \(n\)-MOSFET) with high-\(k\) metal-gate (HK/MG), lightly doped drain (LDD), multiple spacers (multi-spacers), and silicide. Simulation validation shows improvements over other 28 nm devices, with lower static power consumption and notable optimizations in both S.S. (69.8 mV/dec) and DIBL effect (30.5 mV/V).\\

\end{abstract}

\maketitle


\section{Introduction}


\par While the process nodes in the semiconductor industry progress every 2-3 years, the maturation pace of each node lags behind these rapid updates. Thus, it's essential to consider not only Moore's Law but also the maturity of process technologies at each node. Currently in each node still presents unique challenges in the development of specific, gradually maturing process technologies. Scaling at the 28 nm node faces significant challenges, including increased gate current leakage due to quantum tunneling, drain and gate leakage from hot electron effects, and threshold voltage shifting\cite{6839328,9318939}. Nevertheless, the cost-effective nature of 28 nm planar process technology continues to satisfy the requirements of numerous contemporary applications. Given these considerations, there is a pressing need to further refine and optimize 28 nm planar MOSFET technology.

\par This study embarks on a theoretical examination of subthreshold conduction behavior in modern nano devices, proposing innovative planar MOSFET structures designed specifically for the 28 nm process. By incorporating strategic modifications and leveraging mature technologies, the proposed structures aim to enhance device performance. Utilizing Silvaco TCAD tools for simulation, this research demonstrates marked improvements in subthreshold swing and a reduction in drain-induced barrier lowering effects. The simulations also provide validation of the device design through analysis of partial electrical characteristics, establishing a robust foundation for the ongoing development of advanced semiconductor devices.

\section{Structure Design and Process Flow}
\subsection{Theoretical Analysis}
\par 

 As channel length continue to scaling down, quantum effects significantly influence the energy band profile and leakage current, strongly influencing on the the performance of the devices. In their operation, subthreshold conduction is characterized by a current flow when gate source voltage $V_{GS}$ is lower than $V_{t}$. Focusing on the conduction behavior in subthreshold regime, a qualitative analysis is conducted to explore strategies for optimizing two important parameters, S.S. and DIBL. This qualitative analysis also provides guidance for the design and optimization of planar MOSFET structures and processes at the 28 nm node. These efforts aim to reduce the negative impact of quantum effects on the nano-structure, ensuring a balance between device efficiency and performance.

\subsubsection{Analysis of Subthreshold Swing (S.S.)}

In the subthreshold regime, it is commonly assumed that the subthreshold diffusion current is equivalent to the subthreshold current with subthreshold drift current being negligible\cite{sze2021physics,605442}.The subthreshold swing characterizes the gate control capability of a device. 

The subthreshold swing, denoted as $S$, is defined as the increment in gate voltage, $V_{\rm GS}$, necessary to enhance the drain current, $I_{\rm Dsub}$, by an order of magnitude. This parameter is an indicator of the steepness of the $I_{\rm D} - V_{\rm GS}$ curve within the subthreshold regime. A lower value of $S$ indicates stronger gate control, which is reflected in a steeper current curve in the transfer characteristics within the subthreshold region\cite{sze2021physics,hu2010modern}. The expression of S.S. is expressed as:

\begin{equation}
\rm{S.S.} = \frac{dV_{\rm GS}}{d(\log I_{\rm Dsub})} = \frac{n kT}{q \ln 10} = \left[ 1 + \frac{C_{\rm D}(\phi_{\rm S})}{C_{\rm OX}} \right] \frac{kT}{q \ln 10}
\end{equation}

As we can see from the equation, to minimize S.S. effect, modifications can be made to increase the gate oxide capacitance or to reduce the capacitance of the surface inversion layer barrier. Enhancements in gate oxide capacitance can be efficiently realized by employing high-$k$ (H$k$) dielectric materials or by meticulously reducing the thickness of the gate oxide layer \cite{9318939,6839328}. The reduction in the capacitance of the surface inversion layer barrier can be accomplished through careful adjustments to the substrate doping levels.

\subsubsection{Analysis of Drain-induced Barrier Lowering (DIBL)}
The impact of scaling on the band profile is significant, the shorter the channel length, the more significant the drain-induced barrier lowering (DIBL) effect will be. For 28 nm MOS devices, the DIBL effect is significant, lowering both the threshold voltage and the subthreshold swing\cite{hu2010modern}. The precise calculation of DIBL is challenging, thus we consider qualitatively mitigating its effects. For instance, by appropriately enhancing the doping in the drain-body barrier region through Lightly Doped Drain (LDD) technology, the breakdown voltage of this barrier can be increased. Additionally, using a highly doped channel can help to stand more voltage stress, preventing the depletion region from extending towards the source. 
\subsection{Device Structure}
\begin{figure}[h]
\centering
\includegraphics[width=6.9cm, height=5.7cm]{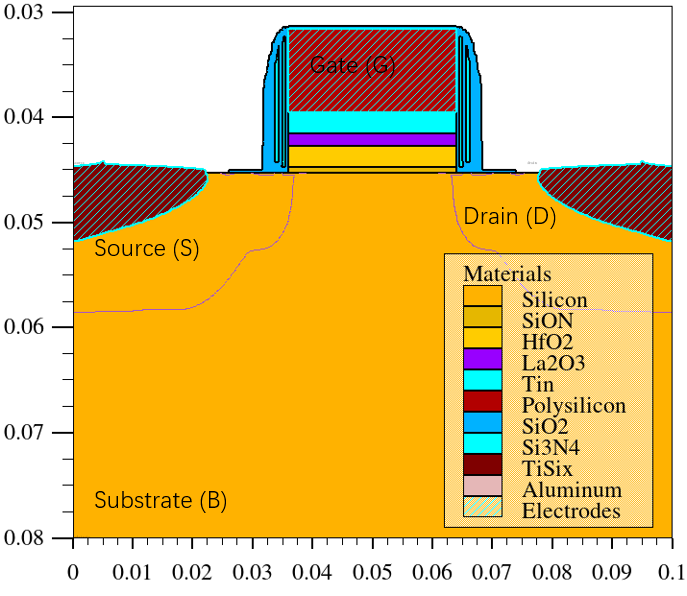}
    \caption{Overall structure of the 28 nm gate length, 100 nm gate width $n$-MOSFET and its materials with unit of $\mu$m.}
    \label{structure}
\end{figure}

As shown in Fig.~\ref{structure}, 28 nm MOSFET is constructed. Detailed parameters about the device such as thickness and doping concentration are listed in Tables I and II. As a result, the overall structure conforms to the typical MOSFET structure with source and drain (S/D), gate, substrate and LDD.

\begin{table}[h]  
  \centering  
\caption{\textbf{Geometric Parameters of Process Simulation}} 
    \begin{tabular}{|c|c|}   
    \hline
    \textbf{Parameters} & \textbf{Size (nm)} \\ 
    \hline
      Gate Length &     28 \\
    \hline
      Gate Width &     100 \\
    \hline
      TiN Thickness & 2 \\
    \hline
     Poly Silicon Thickness & 8 \\
    \hline
    HfO$_2$~Thickness & 2 \\
    \hline
   La$_2$O$_3$~Thickness & 1.2 \\
    \hline
    SiON~Thickness & 0.5 \\
    \hline
    \end{tabular}
\end{table}

\begin{table}[!h]
  \centering
\caption{\textbf{Doping Parameters of Process Simulation}} 
    \begin{tabular}{|c|c|c|c|c|}   
    \hline
    \textbf{Type} & \textbf{Impurities}& \textbf{Dose (cm\(^{-2}\))}& \textbf{Doping (cm\(^{-3}\))} \\ 
    \hline
      Substrate & Boron  & \(1 \times 10^{14}\) & \(3 \times10^{16}\) \\
    \hline
     \(V_{\text{TH}}\) Adjust & Boron  & \(1.2 \times 10^{13}\) & \(5 \times 10^{19}\) \\
    \hline
      LDD & Phosphorous  & \(1 \times 10^{11}\) & \(2 \times 10^{17}\) \\
    \hline
     S/D & Arsenic & \(1.62 \times 10^{14}\) & \(1.2 \times 10^{20}\) \\
    \hline
    \end{tabular}
\end{table}

\subsection{Process Design and Simulation}

\par
\begin{figure}[h]
    \centering
\includegraphics[width=0.4\textwidth]{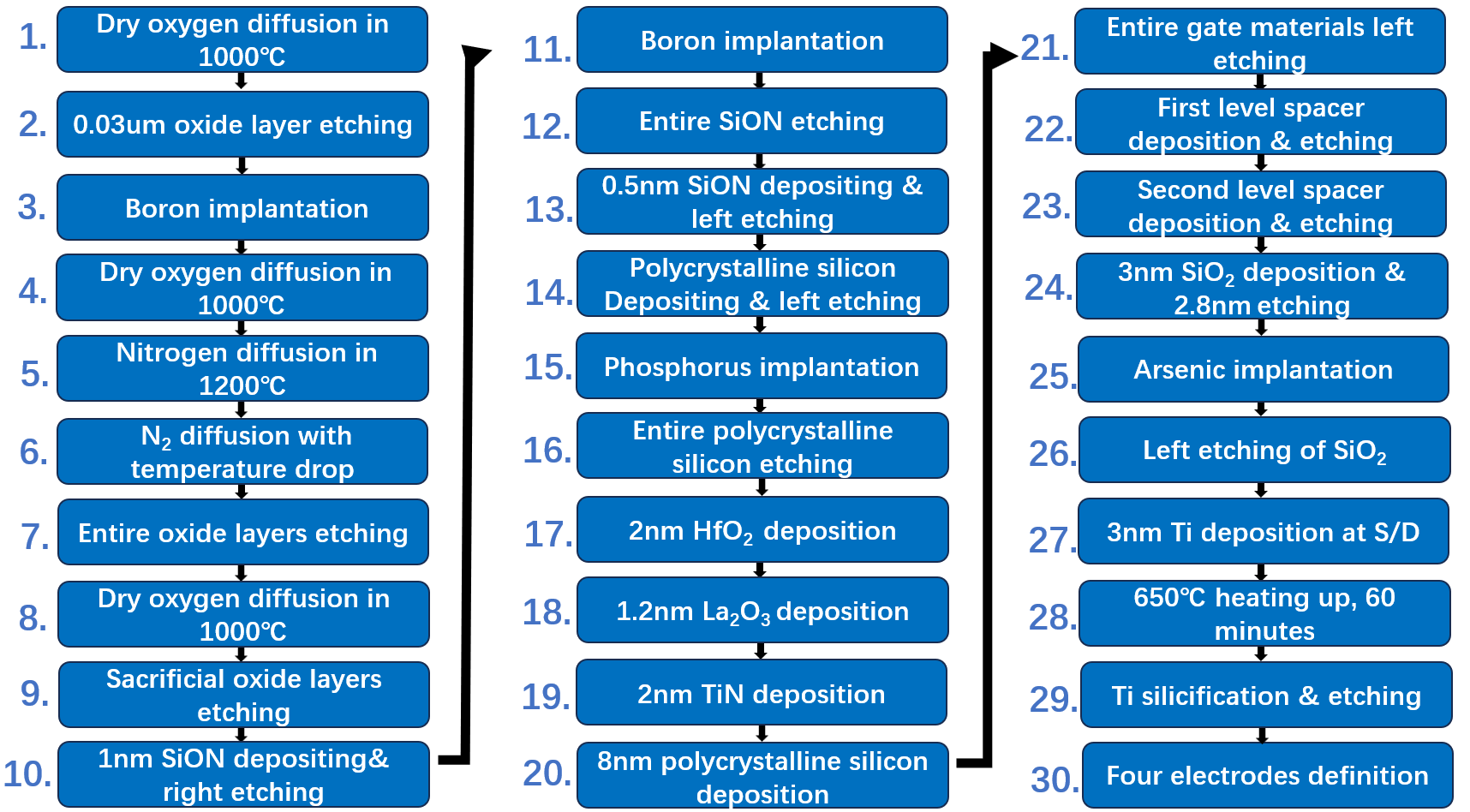}
    \caption{Flow chart of the entire process. }
    \label{Decomposition}
\end{figure}

 The entire process flow is shown in Fig. \ref{Decomposition}.
Basically, the process adopt self-aligned technology and integrate HK/MG, LDD, silicide, and mult-spacers techniques originally.

\begin{figure}
\subfigure{\includegraphics[height=3.8cm,width=3.6cm]{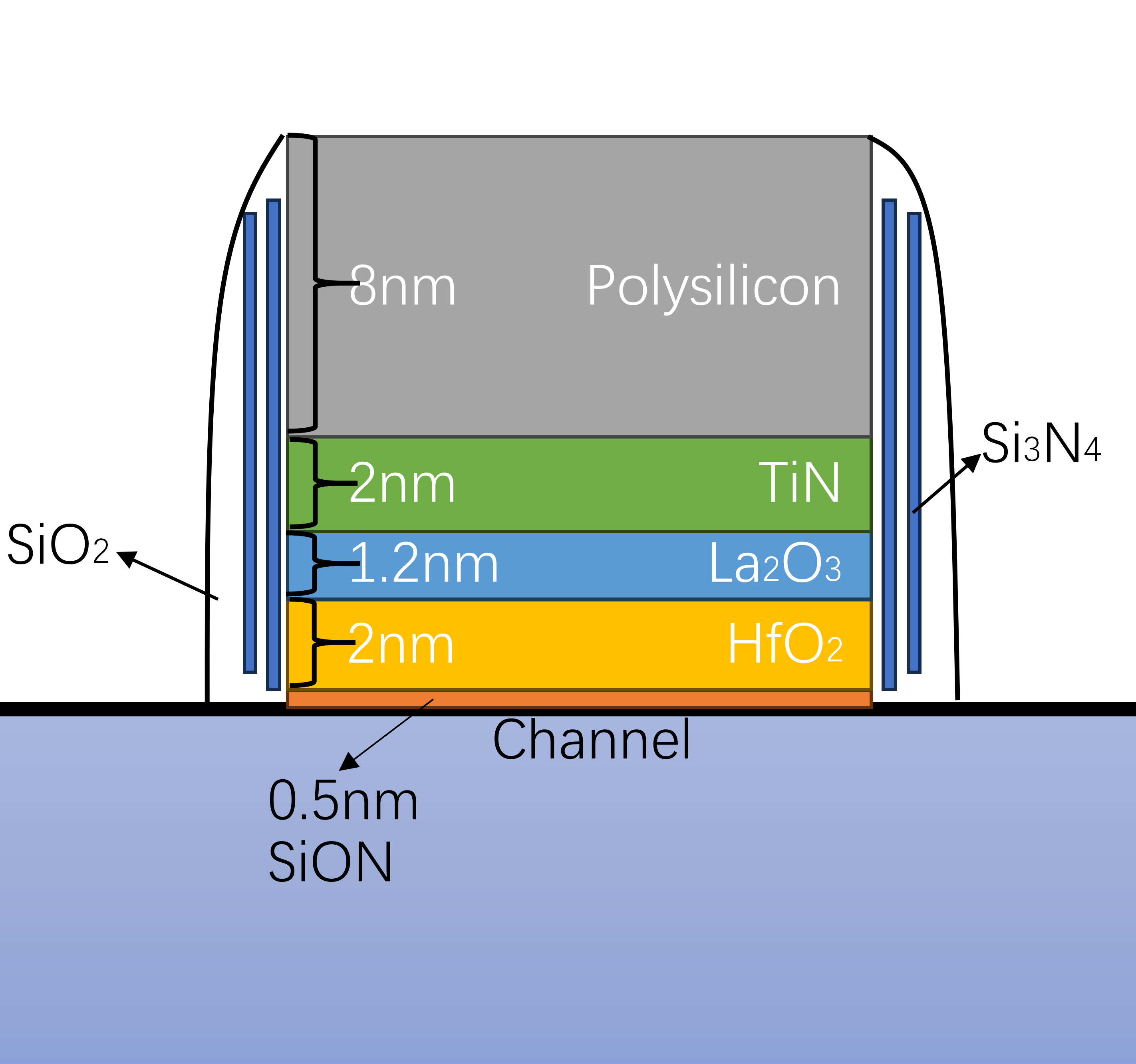}}
\subfigure{\includegraphics[height=4.2cm,width=4.7cm]{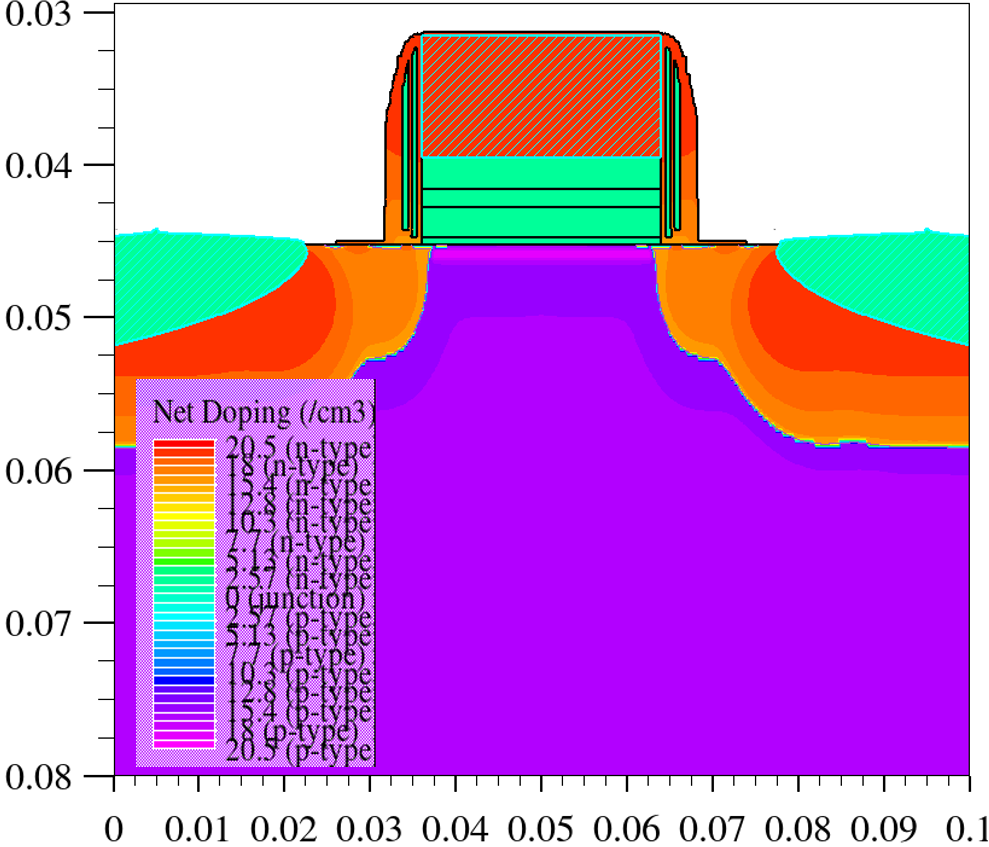}}
   \scriptsize{\textcolor{white}{--------------------------}(a) ~~~~~~~~~~~~~~~~~~~~~~~~~~~~~~~~~~~~~~~~~~~~~~~~~~~(b)}
\caption{(a) Types and thickness of gate materials and 
(b) net doping concentration of the device with unit of $\mu$m.}
\label{doping}
\end{figure}

\par By high-energy (100 keV) boron implantation and Poisson distribution calculation, the fabrication of a retrograde doping well with anti-locking capability and the formation of a $p$-type substrate are completed (steps 1-6 in Fig. \ref{Decomposition}). The process involves high-energy ion implantation (with an implantation energy of 100 keV), positioning the maximum doping concentration, deeper within the well. Based on the chosen Poisson distribution of ion implantation, it is evident that the doping concentration decreases gradually from the maximum concentration towards the device surface, forming an inverted well structure. This retrograde doping well effectively reduces the substrate's parasitic resistance and enhances the device's latch-up immunity.


Subsequently, a sacrificial oxidation layer is used to remove damage and defects from the silicon surface (steps 8-9). 

Then, channel threshold voltage adjustment is achieved by low-energy (0.8 keV) boron surface implantation (steps 10-12).These process begins with surface threshold adjustment ion implantation, increasing the channel surface doping concentration above that of the $p$-type substrate. The fabrication of the lightly doped drain (LDD) structure begins with the deposition of an interlayer dielectric (ILD) between the silicon and gate, utilizing SiON \cite{7959555}, a widely adopted industrial technology known for its excellent stress properties, making it an ideal choice as an interlayer dielectric. Following this, a layer of polysilicon is deposited to act as an implantation mask for the LDD ion implantation, preventing unintended doping in the channel region. The LDD fabrication is carried out using a self-aligned process with phosphorous ion implantation at 1 keV (steps 13-15), effectively reducing the risk of hot-carrier effects by shifting the maximum electric field position away from the current path along the channel surface.

Both the threshold adjustment and LDD ion implantation are surface-level implantations, thus requiring precise control of the implantation energy. Specific ion implantation parameters are listed in Table 2.

After the formation of LDD, a layer-by-layer HKMG gate stack deposition is conducted via the gate-first approach.

 The material and thickness choices are carefully optimized based on process compatibility and device performance. A thin SiON layer (~0.5 nm) is used between the high-$k$ layer and the silicon substrate to mitigate interface defects and maintain electron mobility. HfO$_2$ is selected as the high-k dielectric material for its high dielectric constant and bandgap stability, followed by a 1.2 nm La$_2$O$_3$ layer for flat-band voltage adjustment \cite{10330891}. A TiN metal gate is then applied to address Fermi level pinning and gate depletion issues \cite{10330891,1626448}. The gate stack is completed with a polysilicon layer to form the traditional gate structure (steps 17-21 in Fig. \ref{doping}(a)). After the formation of the LDD, multi-spacers are created using alternating deposition of Si$_3$N$_4$ and SiO$_2$, providing adequate isolation (steps 22-24) \cite{4208596}. Subsequently, arsenic implantation (2 keV) is used to define the S/D regions (step 25), followed by Ti deposition and high-temperature annealing at 650 $^{\circ}\rm{C}$ to form silicide (TiSi$_x$) in the S/D regions, which reduces contact resistance (steps 27-29) \cite{7998177}. The resulting doping profile is shown in Fig. \ref{doping}(b), demonstrating effective control of electrical characteristics through careful engineering of material properties and process steps.

\section{Results and Analysis}
\subsection{S.S. and DIBL Characteristic}
\par
\begin{figure}[h]
\subfigure{\includegraphics[height=4.15cm,width=8.6cm]{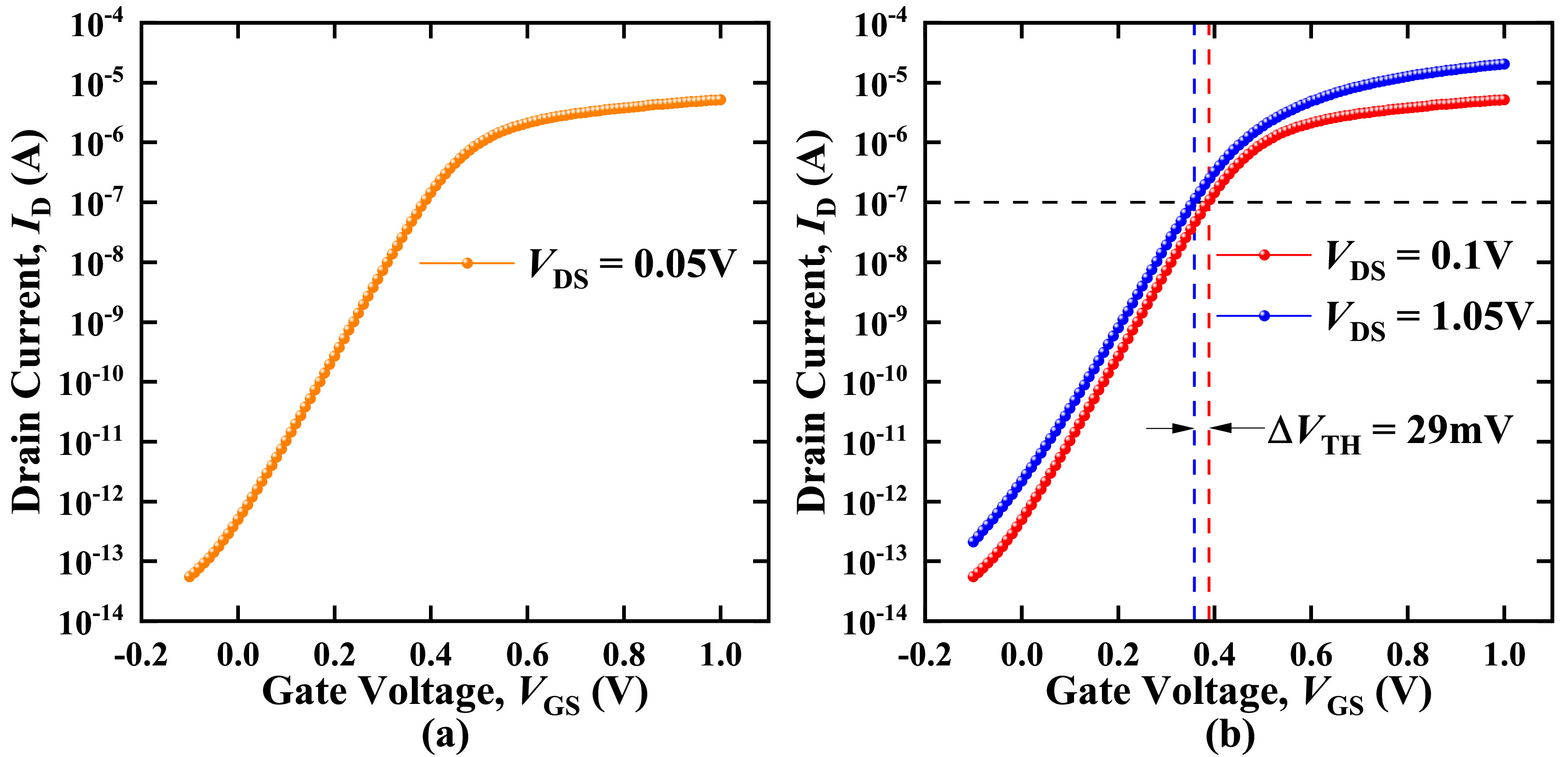}}
\caption{$I_{\rm D}$ versus $V_{\rm GS}$ to indicate (a) the sub-threshold swing and (b) the DIBL effect value. The DIBL is defined as [$\Delta V_{\rm TH}/(1.05\,\text{V} - 0.1\,\text{V})$].}

\label{SS}
\end{figure}

 In the 28 nm node, we take the S.S. at $V_{\rm DS}$ = 0.05 V and $V_{\rm BS}$ = 0 V as the metric to be calculated.
From Fig. \ref{SS}(a), the value of S.S. is  69.8 mV/dec. Similar 28 nm HK/MG MOSFETs were studied in \cite{6839328}, which indicates S.S. of 100-114 mV/dec.\cite{1497095} fabricated 22nm HK/MG CMOS with 95 mV/dec. \cite{6512375} also fabricated 28 mn HK/MG MOSFET with S.S. greater than 75 mV/dec under the same channel length. Assisted by the careful selection of materials, thickness and deposition order in HK/MG,  the device performs 30.2\%, 26.5\% and 6.9\% better than \cite{6839328},\cite{1497095} and \cite{6512375}, respectively.

\par DIBL effect can be quantified through the change of $V_{\rm TH}$  with
$V_{\rm BS}$ equaling to 0 V and $V_{\rm DS}$ ranging from 0.1 V to 1.05 V. $V_{\rm TH}$ is defined as the voltage at which  $I_{\rm D}$ = $10^{-7}~{\rm A}$. As shown in Fig. \ref{SS}(b), the change of $V_{\rm TH}$ is approximately 29 mV. The DIBL value can be calculated and the value is 30.5 mV/V. In the research of \cite{9639002}, the DIBL values are  greater than 50 mV/V for 20 nm Si/InGaAs-FinFET.\cite{1497095} fabricated 22nm HK/MG CMOS with 200 mV/V. \cite{9215330} studied 28 nm CMOS, whose DIBL values is 279 mV/V. With the help of high channel doping concentration and thin gate oxide (shown in Tables I and II), the device performs 85\% , 39\% and 89\% better than \cite{1497095}, \cite{9639002} and \cite{9215330}, respectively.

\subsection{Device Performance Analysis}
Although the device performs well in both S.S. and DIBL values, it is necessary to confirm the device rationality in case of the degradation in other aspects due to excessive pursuit of the two indicators. Physical models of CVT, SRH, BGN, BBT.STD, FLDMOB are considered in the simulation.

\begin{figure}[h]
\subfigure{\includegraphics[height=4.15cm,width=8.6cm]{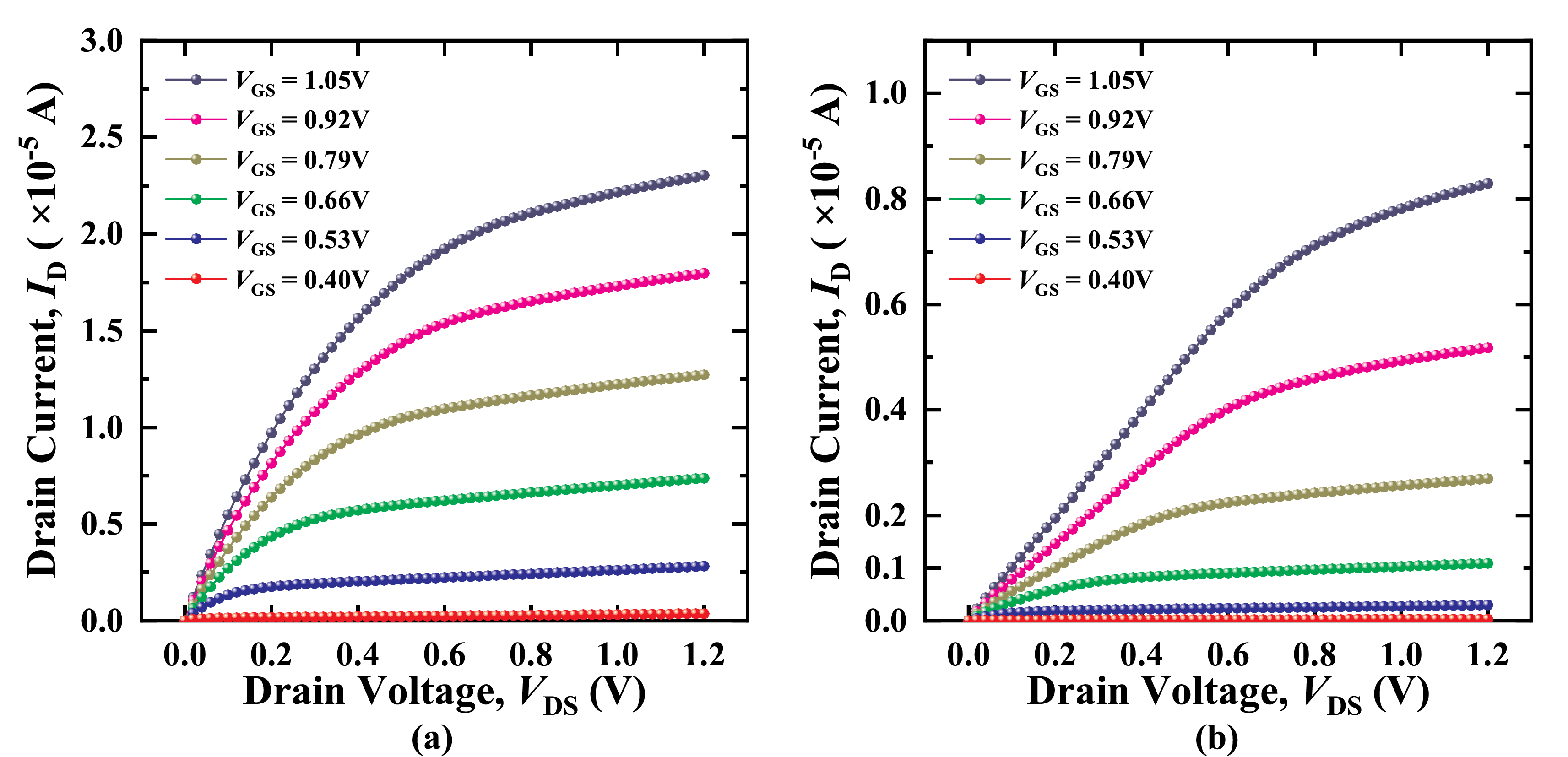}}
\caption{Output characteristic curves with (a) $V_{\rm BS}$ = 0 V and (b) $V_{\rm BS}$ = -1.05 V respectively.}
\label{output}
\end{figure}

\begin{figure}[h]
\subfigure{\includegraphics[height=4.15cm,width=8.6cm]{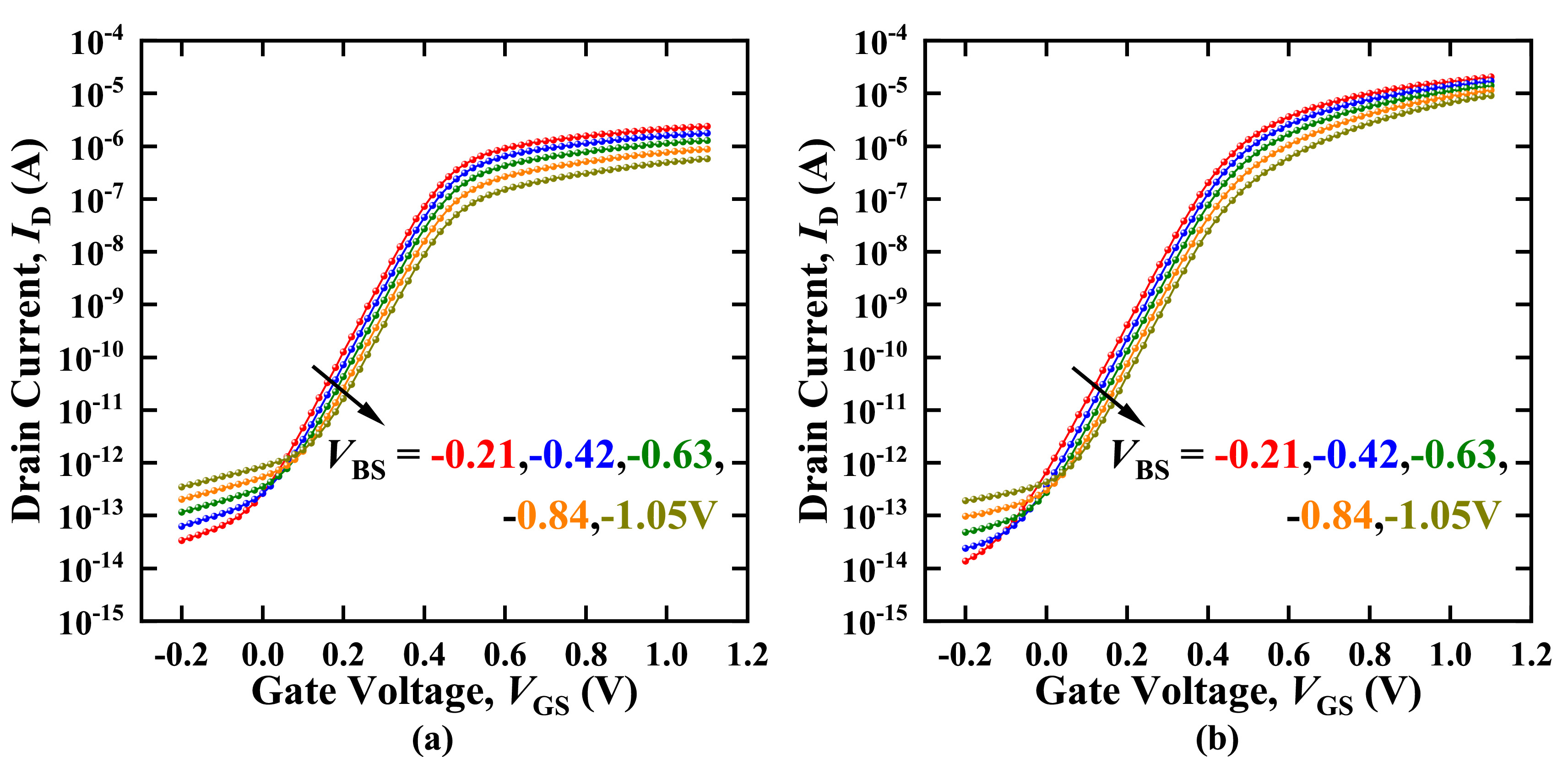}}
\caption{Transfer characteristic curve with (a) $V_{\rm DS}$ = 0.05 V and (b) $V_{\rm DS}$ = 1.05 V. respectively.}
\label{transfer}
\end{figure}

\begin{figure}[h]
\subfigure{\includegraphics[height=4cm,width=8.9cm]{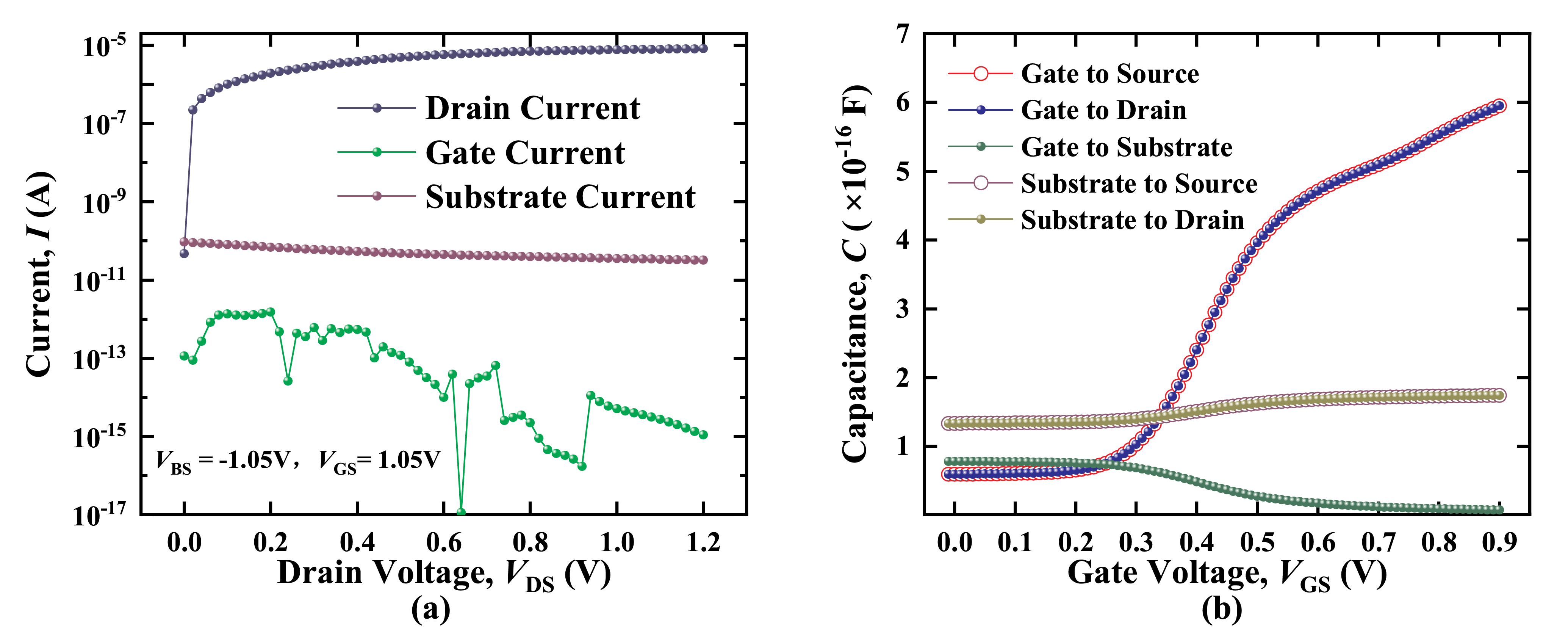}}
\caption{(a) Leakage current of substrate and gate when  $V_{\rm BS}$ =  -1.05 V and $V_{\rm GS}$
= 1.05 V as well as (b) $C$-$V$ characteristic curves under 1MHz.}
\label{leakage}
\end{figure}

\subsubsection{Output and Transfer Characteristic}
\par In Fig. \ref{output}(a) and \ref{output}(b), $V_{\rm BS}$ is 0 V and -1.05 V, respectively. The order of magnitudes (OM) of  $I_{\rm D}$ ($10^{-6}~{\rm A}$) indicates the low power consumption application prospect of the device.

\par In Fig. \ref{transfer}(a) and \ref{transfer}(b), $V_{\rm DS}$ is 0.05 V and 1.05 V, respectively. When $V_{\rm GS}$ = 0 V, OFF current is sufficient low \cite{6512375}, indicating the low static power consumption.

\subsubsection{Leakage Current and C-V Characteristic}
\par In Fig. \ref{leakage}(a), with $V_{\rm BS}$ = -1.05 V and $V_{\rm GS}$
= 1.05 V, the substrate leakage current peaks at OM of $1\times 10^{-10}\rm{A}$ and obviously, the substrate leakage current is sufficiently small  compared to $I_{\rm d}$. The gate leakage current fluctuates due to calculation errors but is generally measurable. Despite of the thin gate oxide, the gate leakage current is limited and reasonable \cite{6839328,9318939} due to the proper adoption of  H$k$ material. The 1MHz $C$-$V$ characteristic shown in Fig. \ref{leakage}(b) demonstrates that the device capacitance is sufficiently small with OM of $10^{-16}~{\rm F}$ \cite{1626448}.

\begin{figure}[h]
\subfigure{\includegraphics[height=4.1cm,width=4.38cm]{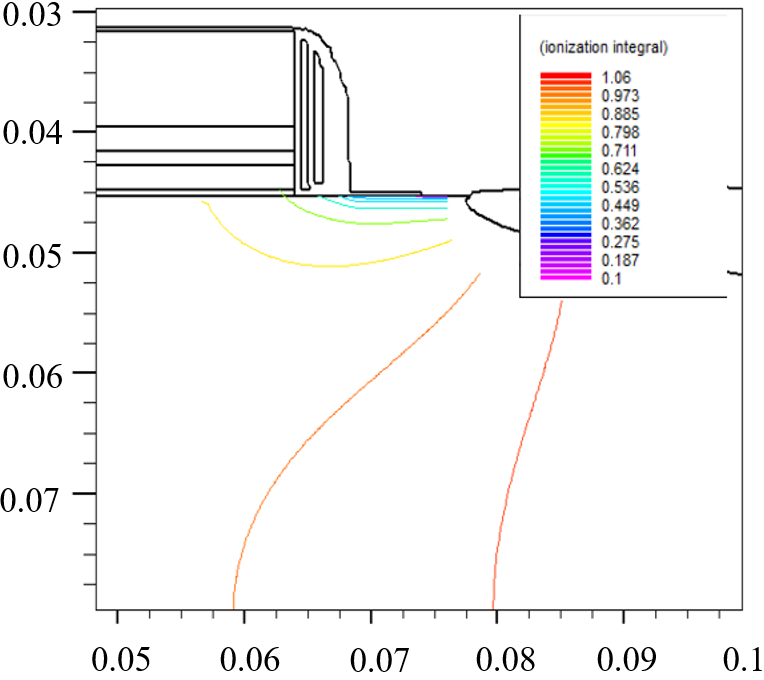}}
\subfigure{\includegraphics[height=4.1cm,width=4.39cm]{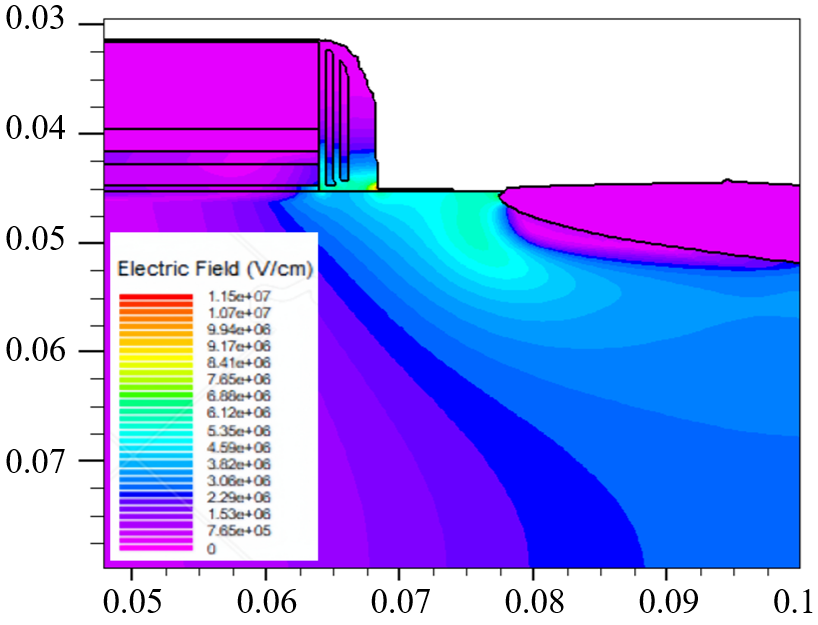}}
\scriptsize{\textcolor{white}{--------------------------}(a) ~~~~~~~~~~~~~~~~~~~~~~~~~~~~~~~~~~~~~~~~~~~~~~~~~~~(b)}
\caption{(a) Ionization integral path and (b) electric field distribution when $V_{\rm DS}$ = 7.35 V with unit of $\mu$m.}
\label{Electric}
\end{figure}

\subsubsection{Electric Field and Breakdown Characteristic}
\par Simulation indicates a breakdown voltage of 7.35 V, suggesting no breakdown under normal conditions ($V_{\rm DS}$ within 0 V to 1.05 V). In Fig. \ref{Electric}(a), it illustrates the ionization integral at 7.35 V, with a peak close to unity along the red line, indicating breakdown occurs. Fig. \ref{Electric}(b) displays the electric field distribution at 7.35 V, with a peak intensity of about  $5.35~{\rm MV/cm}$. This suggests a reasonable electric field distribution and low likelihood of breakdown under normal condition.

\section{Conclusion}

We have successfully designed a 28 nm planar MOSFET that integrates advanced features, including self-aligned technology, LDD, HK/MG, multi-spacers, and silicide. Using Silvaco TCAD simulations, the proposed device has been thoroughly verified in terms of both structure and process. The results demonstrate superior performance, with a 30.2\% reduction in subthreshold swing to 69.8 mV/dec and an 89\% decrease in DIBL to 30.5 mV/V. Overall, the optimized MOSFET design achieves lower leakage current, faster switching speeds, and enhanced power efficiency, making it a robust and a cost effective solution for challenges encountered from 28 nm.


\bibliographystyle{IEEEtran}
\bibliography{ArXiv}

\providecommand{\noopsort}[1]{}\providecommand{\singleletter}[1]{#1}%
\begin{thebibliography}{10}
\providecommand{\url}[1]{#1}
\csname url@samestyle\endcsname
\providecommand{\newblock}{\relax}
\providecommand{\bibinfo}[2]{#2}
\providecommand{\BIBentrySTDinterwordspacing}{\spaceskip=0pt\relax}
\providecommand{\BIBentryALTinterwordstretchfactor}{4}
\providecommand{\BIBentryALTinterwordspacing}{\spaceskip=\fontdimen2\font plus
\BIBentryALTinterwordstretchfactor\fontdimen3\font minus \fontdimen4\font\relax}
\providecommand{\BIBforeignlanguage}[2]{{%
\expandafter\ifx\csname l@#1\endcsname\relax
\typeout{** WARNING: IEEEtran.bst: No hyphenation pattern has been}%
\typeout{** loaded for the language `#1'. Using the pattern for}%
\typeout{** the default language instead.}%
\else
\language=\csname l@#1\endcsname
\fi
#2}}
\providecommand{\BIBdecl}{\relax}
\BIBdecl

\bibitem{6839328}
W.-D. Lee, C.-W. Lian, S.-J. Wang, Y.-H. Yu, O.~Cheng, L.~S. Huang, and M.-C. Wang, ``Electrical quality of 28 nm {HK}/{MG} {MOSFET}s with {PDA} and {DPN} treatment,'' in \emph{2014 International Symposium on Next-Generation Electronics (ISNE)}, 2014, pp. 1--4.

\bibitem{9318939}
D.-C. Zhang, C.-C. Chou, H.-H. Chen, S.-Z. Chen, J.-M. Chen, H.-Y. Bor, W.-H. Lan, and M.-C. Wang, ``Fringe gate leakage of 28 nm {HK/MG} n{MOSFET}s with nitridation treatments,'' in \emph{2020 3rd IEEE International Conference on Knowledge Innovation and Invention(ICKII)}, 2020, pp. 1--3.

\bibitem{sze2021physics}
S.~M. Sze, Y.~Li, and K.~K. Ng, \emph{Physics of Semiconductor Devices}.\hskip 1em plus 0.5em minus 0.4em\relax Hoboken, NJ: John Wiley \& Sons, Inc., 2021.

\bibitem{605442}
E.~Vandamme, P.~Jansen, and L.~Deferm, ``Modeling the subthreshold swing in mosfet's,'' \emph{IEEE Electron Device Letters}, vol.~18, no.~8, pp. 369--371, 1997.

\bibitem{hu2010modern}
C.~Hu, \emph{Modern Semiconductor Devices for Integrated Circuits}.\hskip 1em plus 0.5em minus 0.4em\relax Upper Saddle River, N.J: Prentice Hall, 2010.

\bibitem{7959555}
H.-S. Kim, S.-W. Han, W.-H. Jang, C.-H. Cho, K.-S. Seo, J.~Oh, and H.-Y. Cha, ``Normally-off gan-on-si misfet using pecvd sion gate dielectric,'' \emph{IEEE Electron Device Letters}, vol.~38, no.~8, pp. 1090--1093, 2017.

\bibitem{10330891}
Y.~Wei, J.~Yao, Q.~Zhang, G.~Sang, Y.~Bao, J.~Gao, J.~Li, J.~Luo, and H.~Yin, ``Sub-5-Å {L}a$_2${O}$_3$ in situ dipole technique for large {VFB} modulation with {EOT} reduction and improved interface for {HKMG} technology,'' \emph{IEEE Transactions on Electron Devices}, vol.~71, no.~1, pp. 746--751, 2024.

\bibitem{1626448}
R.~Singanamalla, H.~Yu, G.~Pourtois, I.~Ferain, K.~Anil, S.~Kubicek, T.~Hoffmann, M.~Jurczak, S.~Biesemans, and K.~De~Meyer, ``On the impact of {T}i{N} film thickness variations on the effective work function of poly-{S}i/{T}i{N}/{S}i{O}$_2$ and poly-{S}i/{T}i{N}/{H}f{S}i{ON} gate stacks,'' \emph{IEEE Electron Device Letters}, vol.~27, no.~5, pp. 332--334, 2006.

\bibitem{4208596}
T.~Watanabe, A.~Menjoh, T.~Mochizuki, S.~Shinozaki, and O.~Ozawa, ``A 100Å thick stacked {S}i{O}$_2$/{S}i$_3${N}$_4$/{S}i{O}$_2$ dielectric layer for memory cafacitor,'' in \emph{23rd International Reliability Physics Symposium}, 1985, pp. 18--23.

\bibitem{7998177}
N.~Breil, A.~Carr, T.~Kuratomi, C.~Lavoie, I.-C. Chen, M.~Stolfi, K.~D. Chiu, W.~Wang, H.~Van~Meer, S.~Sharma, R.~Hung, A.~Gelatos, J.~Jordan-Sweet, E.~Levrau, N.~Loubet, R.~Chao, J.~Ye, A.~Ozcan, C.~Surisetty, and M.~Chudzik, ``Highly-selective superconformai {CVD} {T}i silicide process enabling area-enhanced contacts for next-generation {CMOS} architectures,'' in \emph{2017 Symposium on VLSI Technology}, 2017, pp. T216--T217.

\bibitem{1497095}
B.~Doris, D.~Park, K.~Settlemyer, P.~Jamison, D.~Boyd, Y.~Li, J.~Hagan, T.~Staendert, J.~Mezzapelli, D.~Dobuzinsky, B.~Linder, V.~Narayanan, S.~Callegari, E.~Gousev, K.~Guarini, R.~Jammy, and M.~Leong, ``Ultra-thin {SOI} replacement gate {CMOS} with {ALD} {T}a{N}/high-k gate stack,'' in \emph{IEEE VLSI-TSA International Symposium on VLSI Technology, 2005. (VLSI-TSA-Tech).}, 2005, pp. 101--102.

\bibitem{6512375}
M.-C. Wang, C.-K. Du, M.-R. Peng, S.-J. Wang, S.-Y. Chen, C.-H. Liu, O.~Cheng, L.~S. Huang, and S.~C. Lee, ``Trend of subthreshold swing with {DPN} process for 28 nm {N/PMOSFET}s,'' in \emph{2013 International Symposium on Next-Generation Electronics}, 2013, pp. 389--392.

\bibitem{9639002}
S.-E. Huang, W.-X. You, and P.~Su, ``Mitigating {DIBL} and short-channel effects for {III-V} {FinFET}s with negative-capacitance effects,'' \emph{IEEE Journal of the Electron Devices Society}, vol.~10, pp. 65--71, 2022.

\bibitem{9215330}
A.~Kumar, M.~Pattanaik, P.~Srivastava, and K.~K. Jha, ``Dual metal hetero dielectric {GAAFET} based energy efficient digital circuits,'' in \emph{2020 International Conference on Smart Electronics and Communication (ICOSEC)}, 2020, pp. 1206--1209.

\end{thebibliography}

\end{document}